\documentclass[prn]{aa}  
\usepackage{graphicx,natbib}
\bibpunct{(}{)}{;}{a}{}{,} 
\begin{document}

\title{The colour of the narrow line Sy1-blazar 0324+3410}

   \author{S. Ant\'on
          \inst{1}
          \and
          I.W.A. Browne 
          \inst{2}
          \and
          M.J. March\~a
          \inst{3}
          }

   \offprints{santon@fc.ul.pt}

   \institute{SIM-IDL, Faculdade de Ci\^encias da Universidade de Lisboa\\
              Campo Grande, C8, 1749-016 Lisboa, Portugal
              \email{santon@fc.ul.pt}
         \and University of Manchester, Jodrell Bank Centre for Astrophysics,\\
               Alan 
              Turing Building, Oxford Road, Manchester, M13 9PL, U.K.
          \and Faculdade de Ci\^encias da Universidade de Lisboa\\
              Campo Grande, C8, 1749-016 Lisboa, Portugal
             }

   \date{Received October 2007 ; accepted --- }

   \abstract
   {}
   {
We investigate the properties of the host galaxy of the 
    blazar J0324+3410 (B2 0321+33) by the analysis of B and R images obtained with the NOT 
under good photometric conditions.
   }
   {The galaxy was studied using different methods: Sersic model fitting, 
unsharp-masked images, B-R image and B-R profile analysis.}
   {
The images show that the host galaxy has a ring-like
morphology. The B-R colour image reveals two bluish zones: one that
coincides with the nuclear region, interpreted as the signature of
emission related to the active nucleus, the other zone is extended and
is located in the host ring-structure. We discuss the hypothesis that
the later is thermal emission from a burst of star formation triggered
by an interacting/merging process.
   }

   \keywords{
galaxies: galaxies: active - galaxies: peculiar - galaxies: photometry.
               }

   \maketitle
\section{Introduction}
The blazar J0324+3410 (B2 B0321+33) is a relatively strong radio source, with 
a flat spectrum stretching up to high frequencies, the spectral energy 
distribution showing the synchrotron peak at the optical band (e.g. Ant\'on et
 al. 2004). Recently Zhou et al. (2007, hereafter Zhou et al.) presented a 
through  analysis of its properties and concluded that it is a blazar in a 
narrow line Seyfert 1 galaxy (NLS1). In terms of the host galaxy morphology, 
they noted a ring like structure and suggested that it might be an asymmetric 
spiral arm. \\ 
The nature of NLS1 objects is not well understood, particularly the narrowness
of the permitted lines. There are results suggesting
 that NLS1s are similar to the Broad emission Line Seyfert 1 (BLS1) but with
 a  face-on disk-like broad emission line region (BLR ;McLure \& Dunlop 2002), 
or with a partly obscured BLR (Smith et al. 2002). But there are also works 
that argue that NLS1s may be Seyfert galaxies in their early stage of 
evolution, with low mass black holes, but high accretion rate (e.g. Boroson 
2002, Mathur 2000). Most of the NLS1 objects are radio quiet  (Zhou et al.) 
and very rarely  show ring structures (Ohta et al 2007). It is
therefore interesting that J0324+3410 is a radio-loud object and its host shows 
a ring-like structure.  
Ring galaxies belong to the class of galaxies that suffer a bulls-eye collision
with another galaxy (see the review of Appleton \& Struck-Marcell, 1996).
It is of obvious interest the case of a galaxy showing
disturbed morphology that  
harbours an active galactic nuclei, as the role of interactions/mergers in 
triggering AGN is on debate (e.g. Alonso et al. 2007), plus the fact 
that according to some NLS1 models (see above), the black hole of J0324+3410 
might be in a growing  phase. \\
Here we present B and R images of J0324+3410 that were obtained with the
2.6~m Nordic Optical Telescope (NOT) during a project designed to
examine the host properties of the 200-mJy sample (Ant\'on 2000, Ant\'on
 et al. 2002),
 a low luminosity radio-loud sample, of which J0324+3410
is a member. In that sample we found that most of the objects had very
similar large-scale properties, like host-galaxy type and
environment. We also found few objects with interesting features,
J0324+3410 being one of them. Here we analyse its B-R colour image and
its radial colour profile, and we discuss the origin of the ring-like
structure.  Throughout this paper it is assumed H$_o$ = 71 kms$^{-1}$
Mpc$^{-1}$, $\Omega_M$=0.27, $\Omega_\lambda$=0.73.

\section{The observations and data reduction}
In August of 1997 we obtained B and R images\footnotemark\footnotetext{B and R 
filter characteristics, respectively: central wavelength of 4280 and 6310 \AA, 
FWHM of 1050 and 1210 \AA, maximum transmittance of 62\% and 71\%} of 
J0324+3410 with the NOT
using the HiRAC camera which has a 2k Loral CCD, giving a field view
of $3.7\times3.7$ arcmin$^2$ and a pixel scale of 0.11 arcsec. The exposure 
time for B and R filter was of  720 and 480 seconds respectively.
 The conditions were photometric throughout the observations with seeing
$\sim$ 0.7 arcsec.  The data reduction was performed with 
IRAF, following the standard procedures: the images were debiased, trimmed and 
flatfielded. Aperture photometry was performed for the standard stars, 
and  standard calibrations were computed to
obtain the zero point magnitudes: ${\mbox B}_{st}= 0.54 + {\mbox B}_{ins} - 0.27 \cdot X$ and ${\mbox R}_{st}= 0.22 + {\mbox R}_{ins} - 0.11 \cdot X $, were X 
represents the air mass, ``st'' means the calibrated magnitude, ``ins'' means 
the observed magnitude. The statistical 
error  of the calibrations is  RMS=0.03 for B magnitudes  and RMS=0.02 for 
R magnitudes.  A Point Spread Function (PSF) model was derived independently 
from each image, using {\bf daophot} tasks. As many stars as
 possible were included, even the very faint ones, in order 
to achieve a higher PSF dynamic range. 
The colour image B-R was obtained as B-R $\sim$ $\frac{\mbox{(R - R}_{\mbox {\tiny backg}}{\mbox )}/{\mbox t}_{\mbox{\tiny R}}}{\mbox{(B - B}_{\mbox{\tiny backg}}{\mbox )}/{\mbox t}_{\mbox {\tiny B}}}$, where ``backg'' and ``t'' refer to background counts 
and exposure time  respectively. In order to check that the colour features 
were not artifacts we examined the inner structure of objects that we  were 
confident to  be stars. The colours of all the field (non-saturated) stars did
 cancel well. 
Isophotal profiles were obtained through the IRAF routine {\bf ellipse},
which is based on the method developed by Jedrzejewski (1987). From the  radial 
isophotal profiles effective radii r$_e$ were estimated (semimajor axis of the 
isophote enclosing half of the total light): r$_e$(B) = 1.2 kpc and r$_e$(R) = 
1.7 kpc. The colour profile was obtained by subtracting the isophotal profiles in the two bands. The profile is determined up to a maximum radius which is the smallest of the two radii corresponding to the sky brightness of 3$\times$ $\sigma_{sky}$ in each band.
We note that both B-R image and profile were built with a shifted and degraded R image:   the R image was shifted in position to get a 
perfect matching between the frames. Galactic stars were used to obtain the
registration of the frames, then, the shifted R image was smoothed
with a Gaussian filter in order to match the slightly worse resolution
of B image; the task {\bf gauss}, which convolves an image with an
elliptical Gauss function, was used for this purpose.
\begin{figure*}
\centering
\includegraphics[height=6cm]{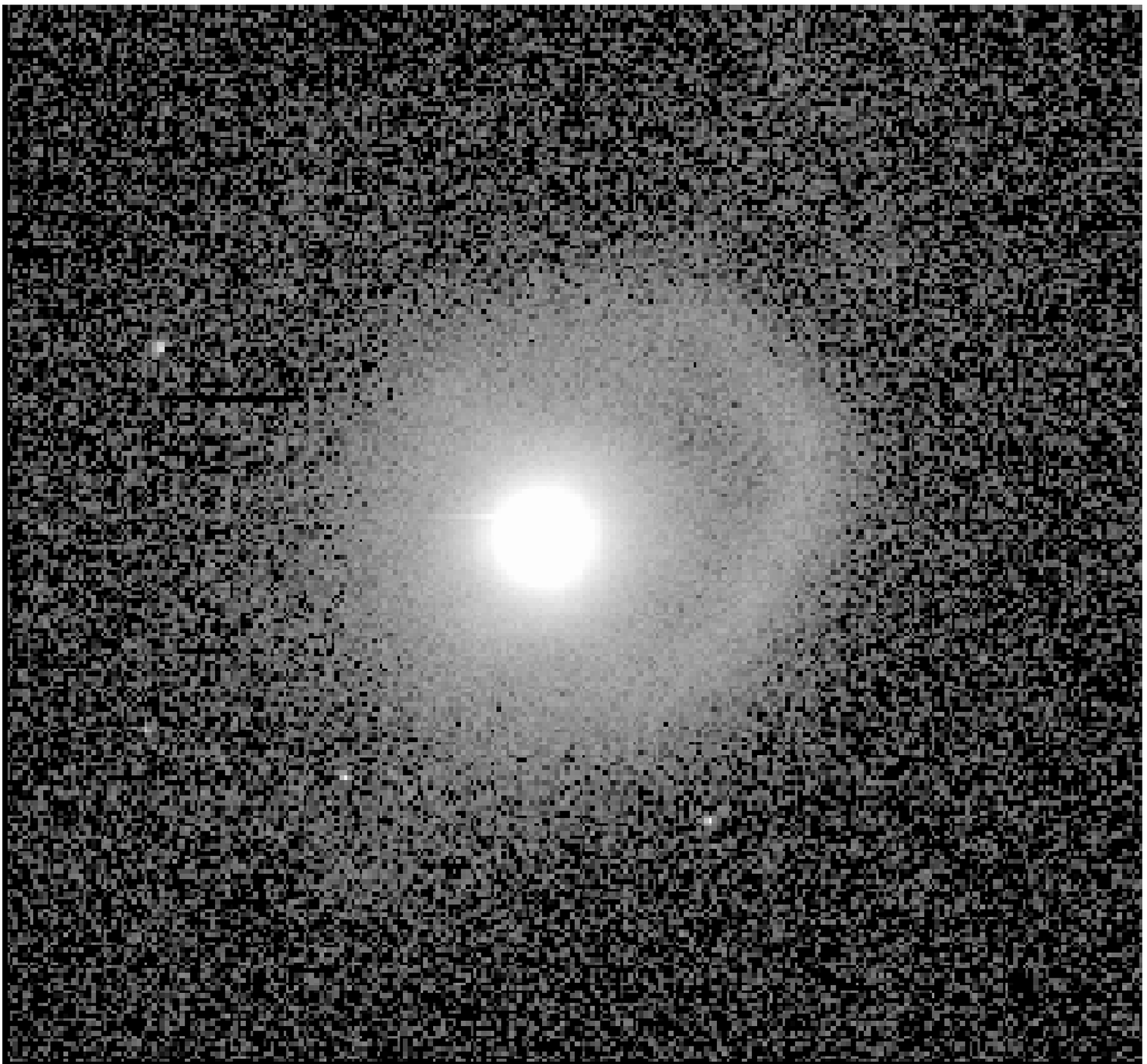}
\includegraphics[height=6cm]{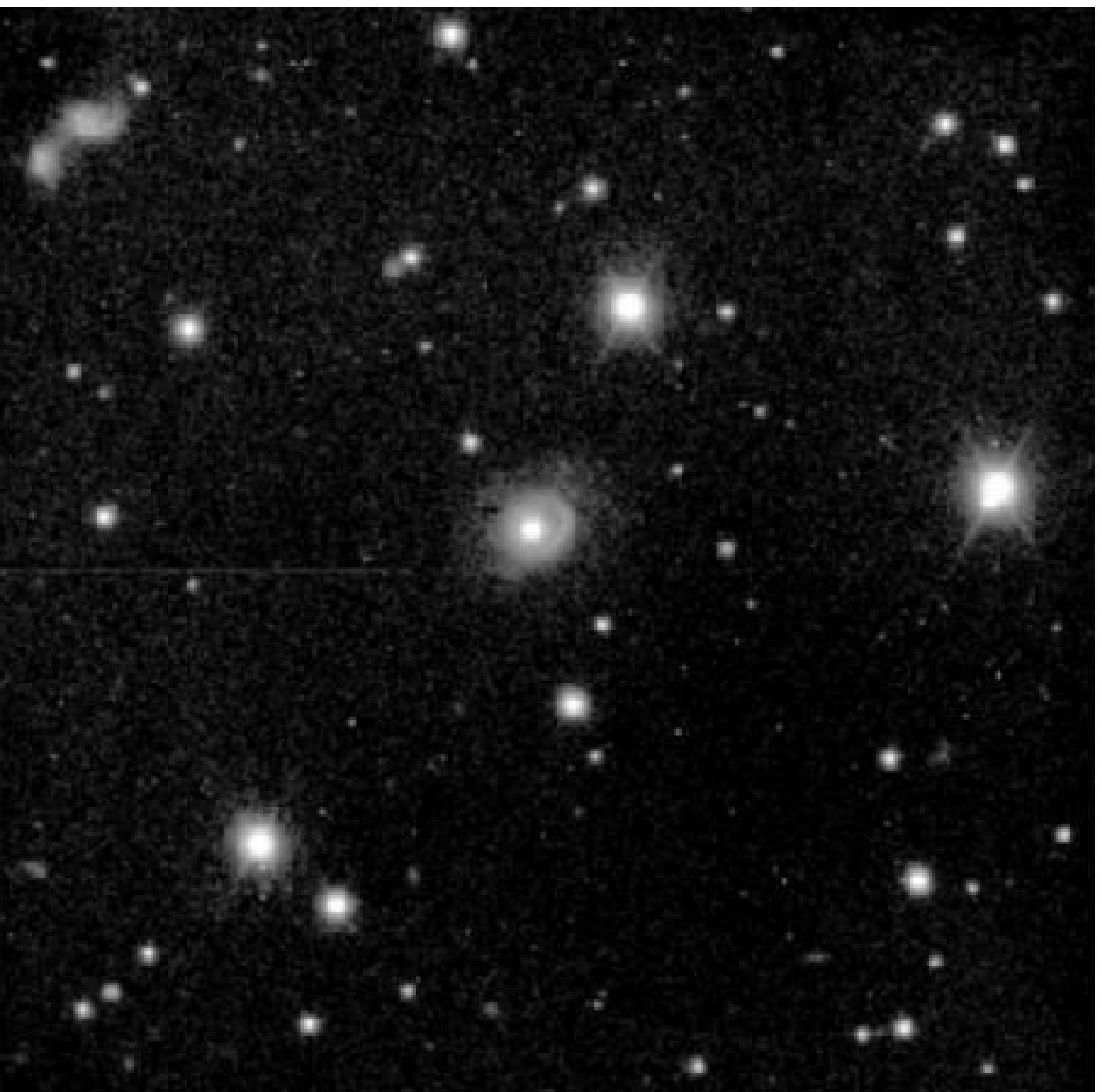}
\includegraphics[height=6cm]{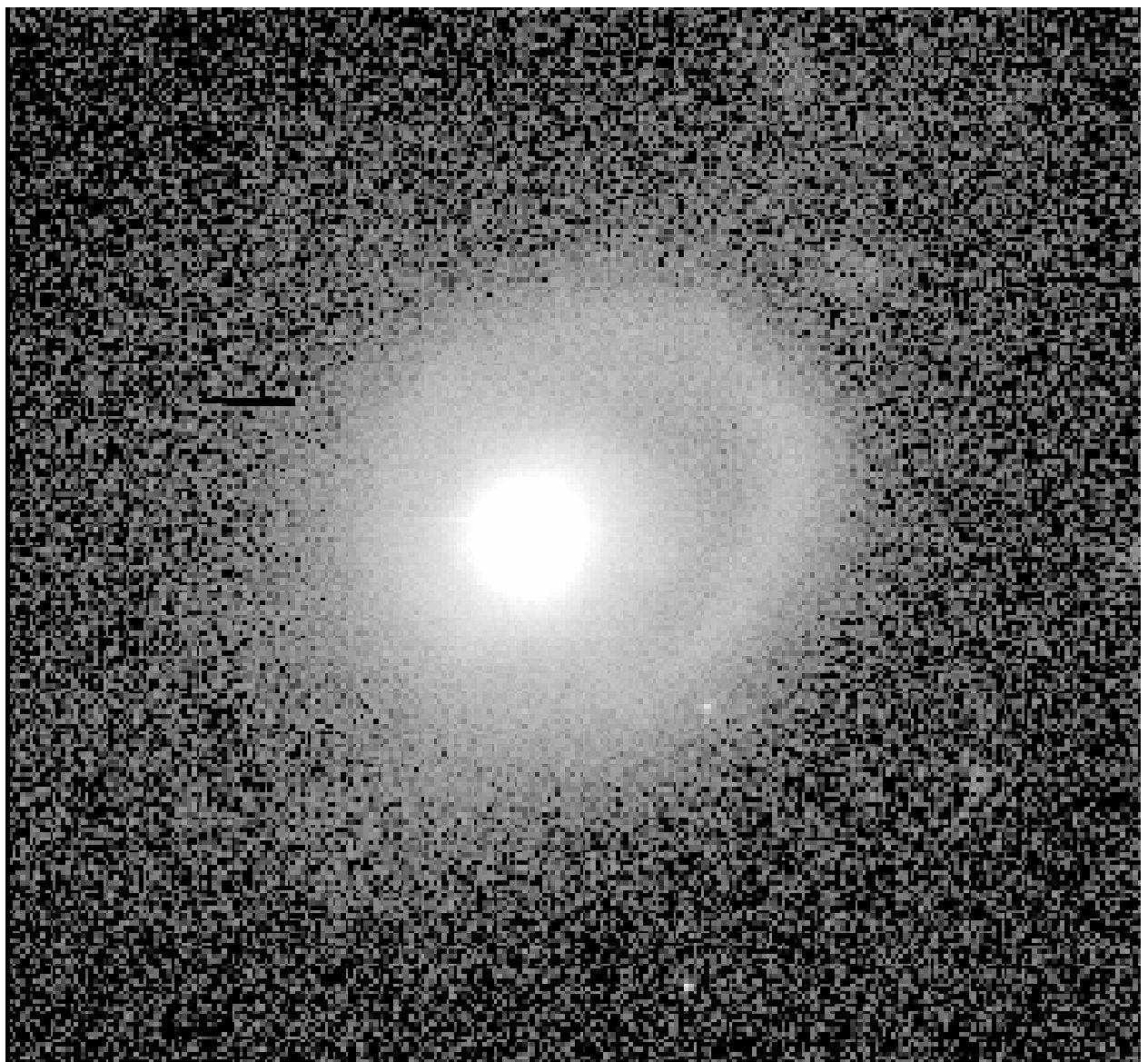}
\includegraphics[height=6cm]{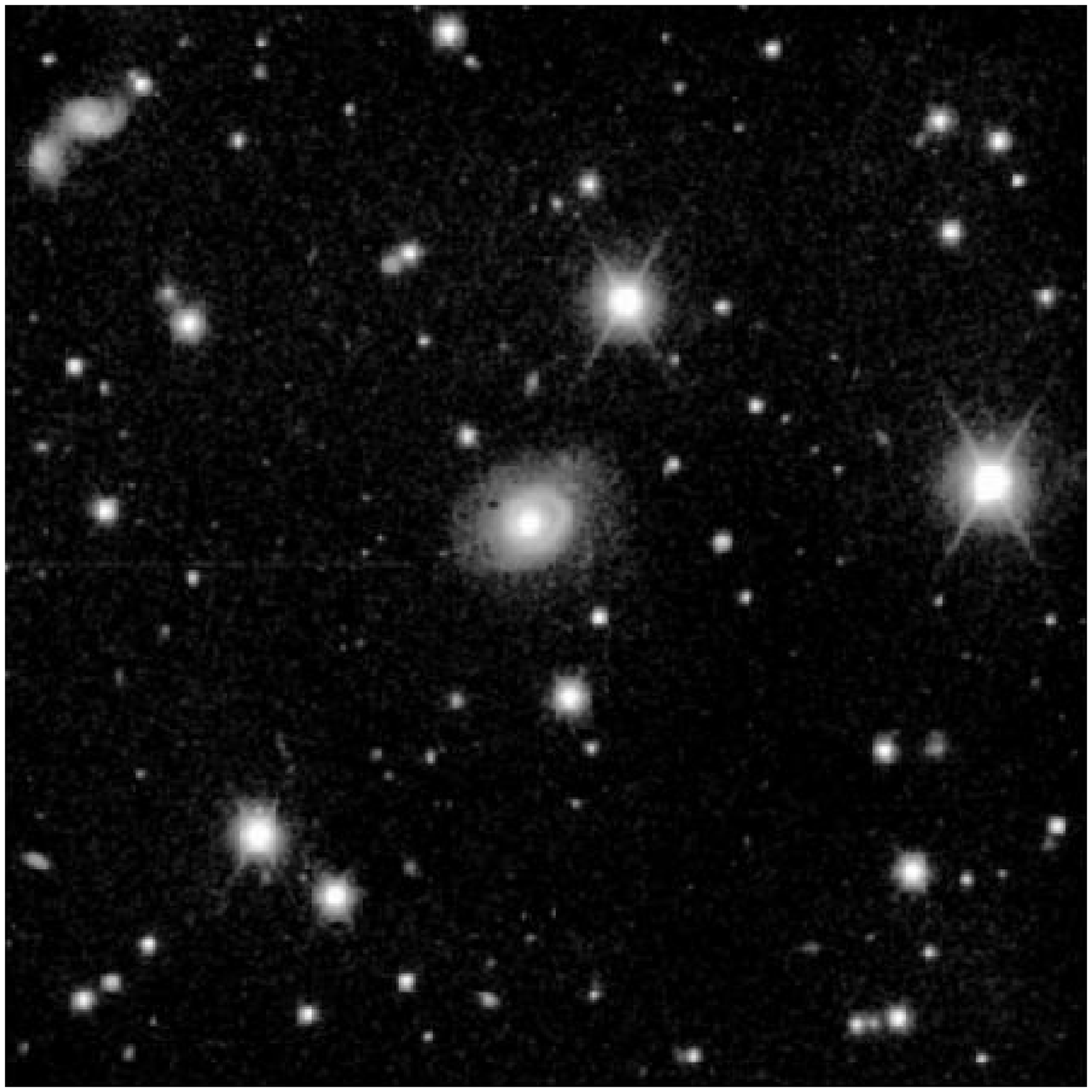}
\caption{0324+3410: {\bf top left -} B image of the object 
(field of view of $29''\times 30''$) and {\bf top right -} the object 
is centred in a field of view of $3.3'\times 3.3'$,
North is up and East is left; {\bf bottom  -} as before, in R band. } 
\label{br}
\end{figure*}
\noindent In order to highlight any features in the host galaxy we
analysed the images through 2-D fitting, using two different methods: 
a) we fitted a Sersic
model to each image, and analysed the image after that model was
subtracted, and b) we created an unsharp-masked image.  The Sersic
model was constructed based on GALFIT package (Peng et al. 2002). The best
model fitting was obtained with a three component Sersic model with
indexes of 8, 4 (Vaucouleur profile) and 0.1 for both B and R
bands. The unsharp-masked image was built by dividing each image by that
image convolved with a Gaussian function of $\sigma$=8 pixel -- the
task {\bf gauss}, was used for this purpose.

\begin{figure*}
\centering
\includegraphics[height=5cm]{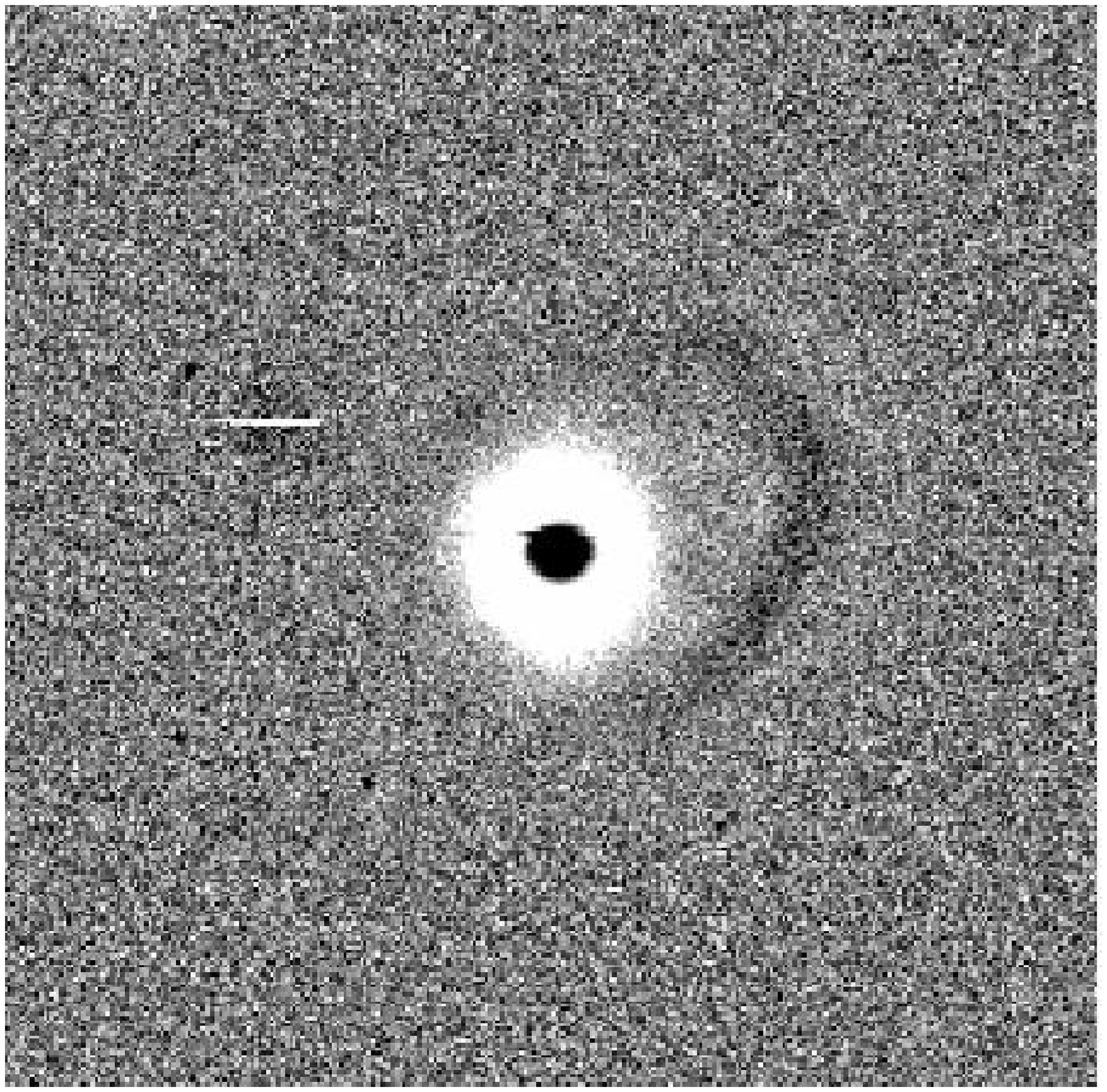}
\includegraphics[height=5cm]{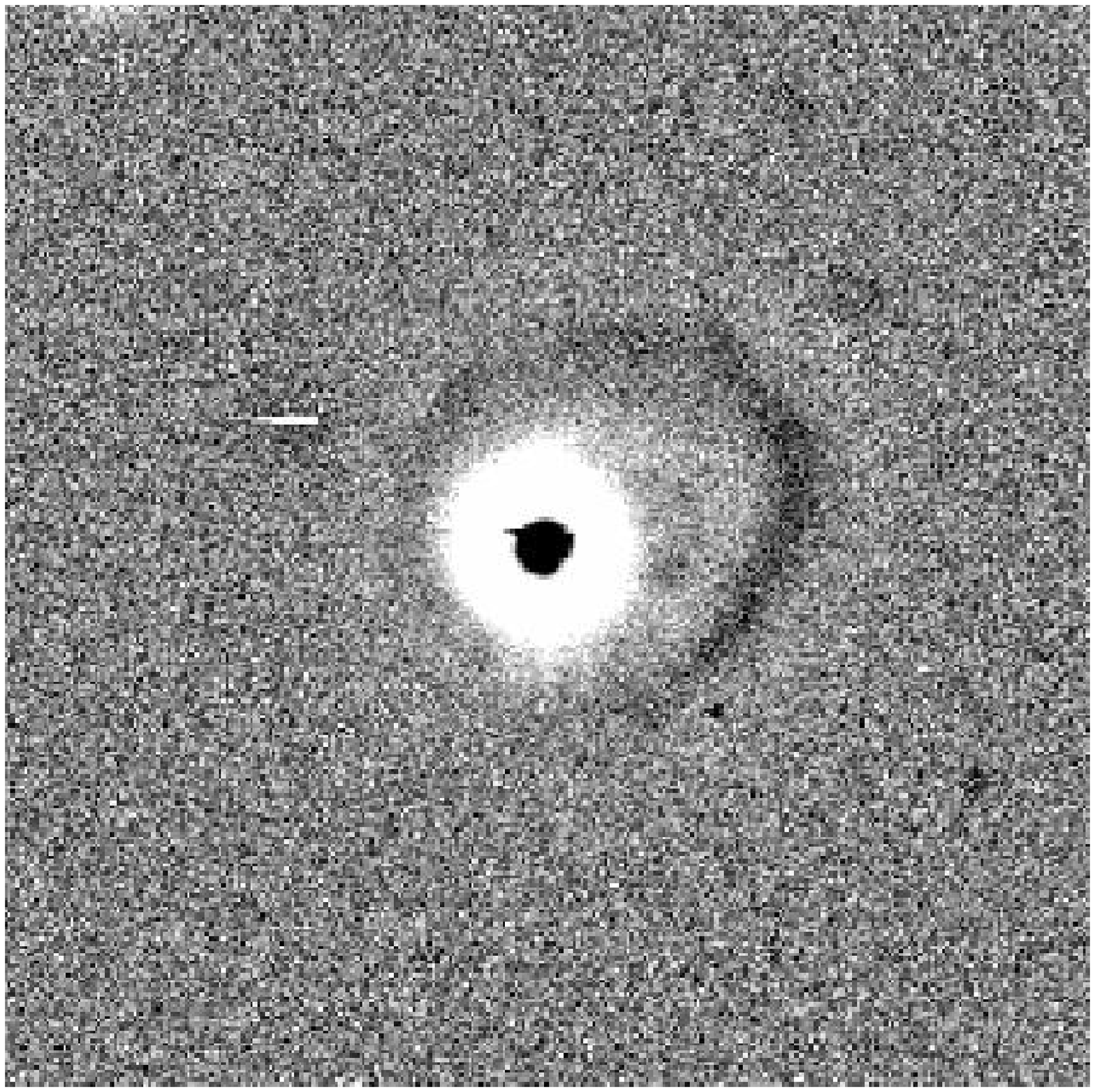}
\caption{0324+341: {\bf left -} unsharp-masked B image,  {\bf right -}
unsharp-masked R image.  Darker/lighter pixels correspond to 
brighter/dimmer regions respectively.}
\label{unsharp}
\end{figure*}

\begin{figure*}
\centering
\includegraphics[height=5cm]{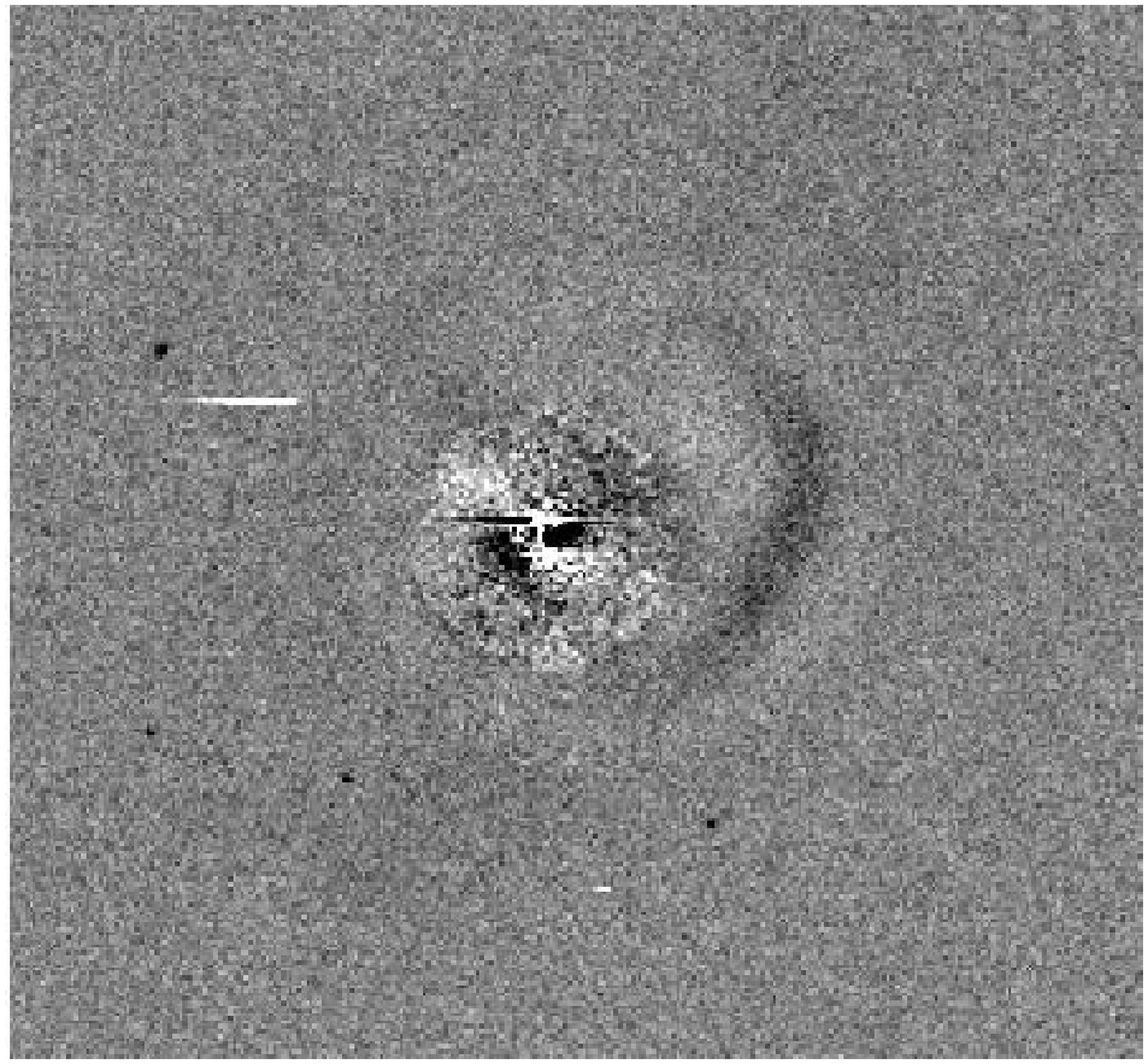}
\includegraphics[height=5cm]{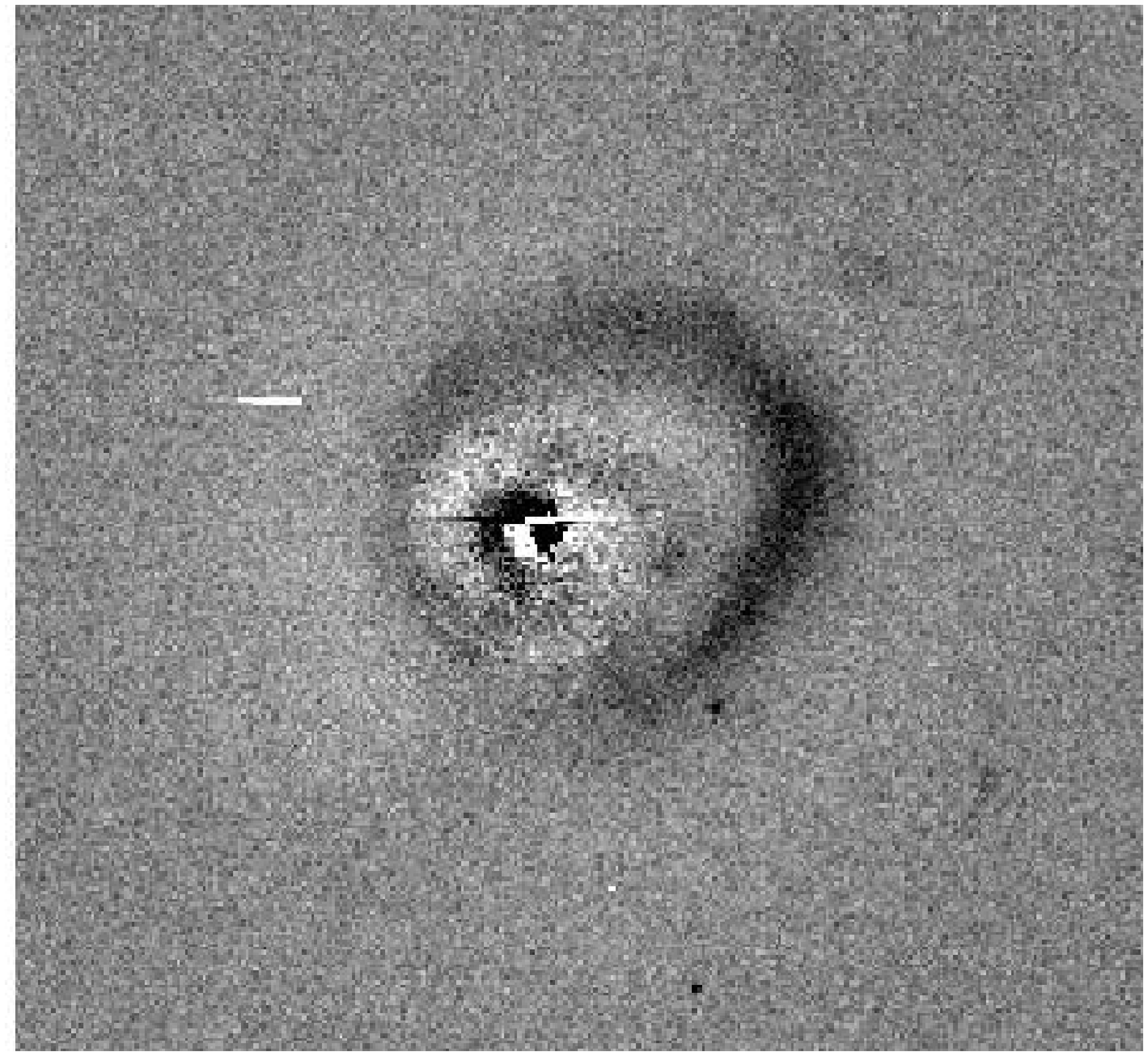}
\caption{0324+341: residual images obtained by subtracting a Sersic profile to 
the {\bf left -} B image, and  {\bf right -} R image. Darker/lighter pixels correspond to 
brighter/dimmer regions respectively.}
\label{residual}
\end{figure*}

\section{Discussion}

 B and R images of J0324+3410, plotted with a logarithmic scale, are
presented in Figure \ref{br}. We also present the same images but with
a larger field of view. Both B and R images show a faint outer
asymmetric disk-like emission. It is also quite clear that the host
galaxy of J0324+3410 has a peculiar morphology of approximately 15$''$
projected diameter ($\sim$ 17 kpc), which is consistent with Zhou et
al.  findings. These authors present an HST image taken with WFPC2
with the F702W filter, and they suggest that the host galaxy looks
like a one-armed spiral. We draw attention to its ring-like appearance
and suggest that it resembles the ring found in the inner region of
the interacting galaxy Arp 10 (Charmandaris \& Appleton 1996,
hereafter CA 96). The ring-like feature is clearly shown in the
unsharp-masked B and R images, Figure~\ref{unsharp}, and also in the
residual images obtained by subtracting the Sersic model (as described
in the previous section) to the galaxy, Figure~\ref{residual}. Note
that in order to best display the ring-shape feature the images are
displayed with inverted code: darker/lighter pixels correspond to
brighter/fainter regions respectively.\\
\begin{figure*}
\centering
\includegraphics[width=7cm,height=5.5cm]{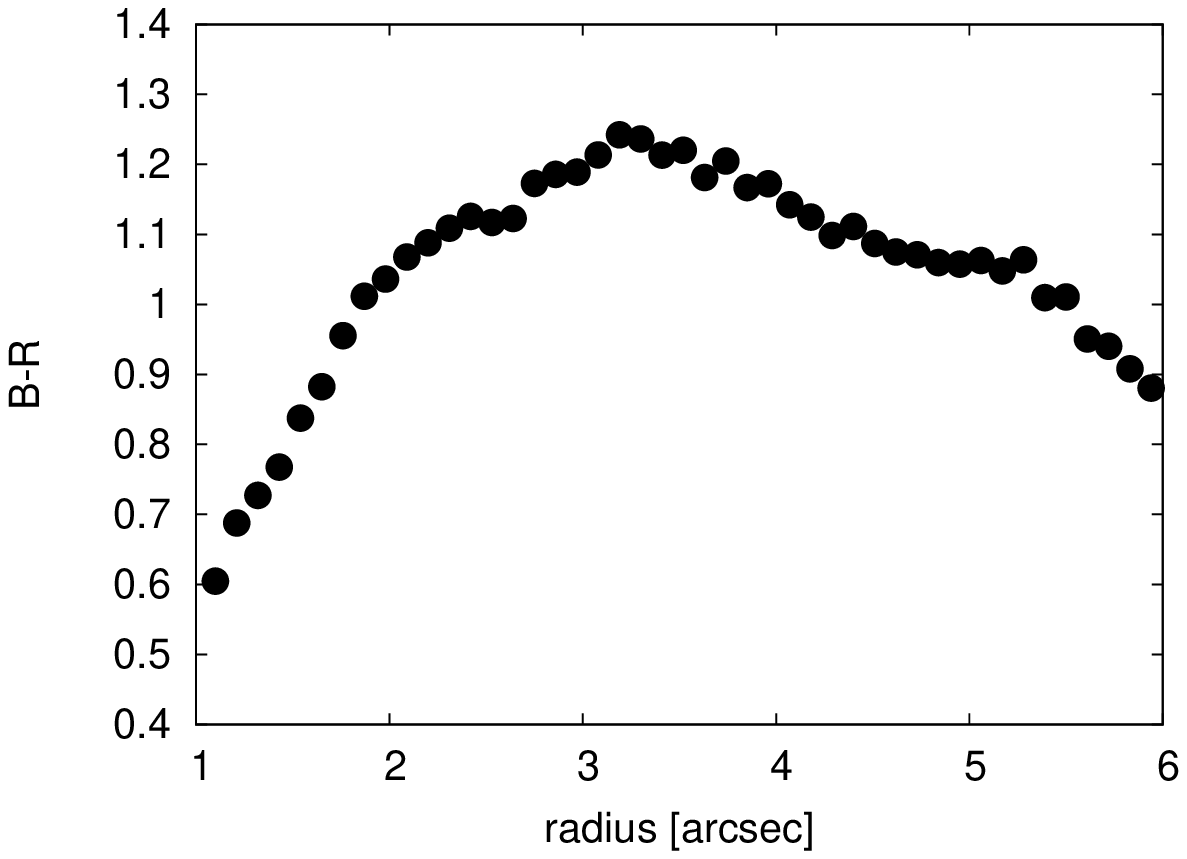}
\includegraphics[height=5.5cm]{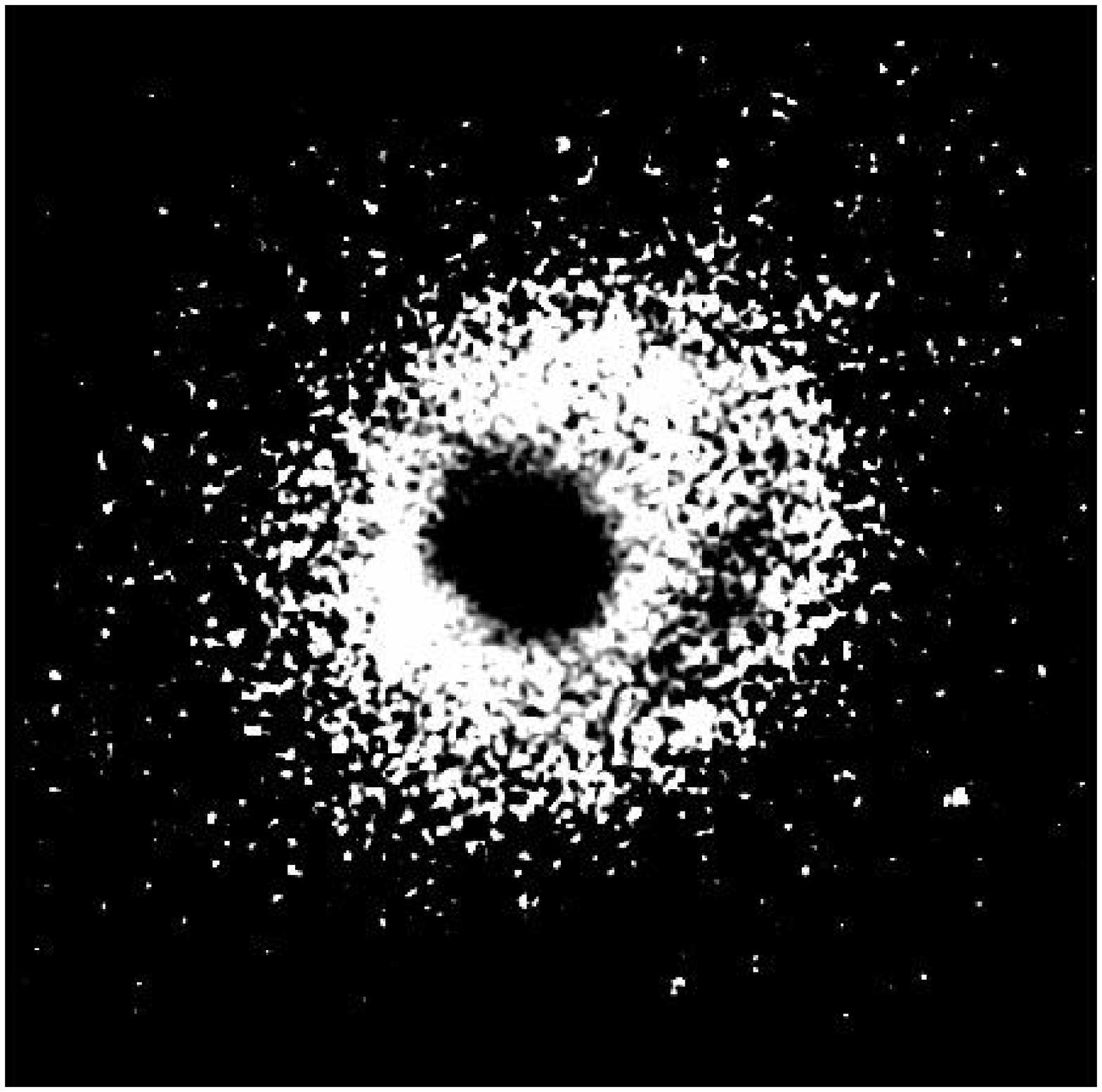}
\caption{{\bf left -} B-R colour profile {\bf right -}  B-R colour image.
North is up and East is left, and the field of view is $28^{''}\times 28^{''}$}
\label{bmrprofile}
\end{figure*}
\noindent Colour images can be useful
for tracing gradients in the stellar population, reddening by dust, or
even scattering in the nuclear regions of AGNs (see Kotilainen \&
Ward, 1997).  The B-R image and profile are presented in Figure
\ref{bmrprofile}. Darker pixels represent bluer colours and lighter
pixels represent redder colours in the colour image. The later was
smoothed with a Gaussian function, with a kernel radius of 3, and the
contrast was adjusted in a way that highlights the features discussed
in the following.  If we fit J0324+3410 profile, Figure~\ref{bmrprofile}, 
between 3$<$r(arcsec)$<$ 6, we obtain $ \Delta{\mbox{\rm
(B-R)}}$/$\Delta(\log\mbox r) $=$-0.6\pm 0.03$ (magnitudes per
arcsec$^2$ per decade in radius), i.e. the object becomes bluer in the
outer regions. Elliptical galaxies in general, and low luminosity
radio galaxies in particular, show a systematic trend to become bluer
at the outer regions. For example radial colours $\langle
\Delta{\mbox{\rm (B-R)}}$/$\Delta(\log\mbox r) \rangle$ of $-0.20\pm
0.04$ and $-0.16\pm 0.17$ are found among radio galaxies (Govoni et
al. 2000, Mahabal et al. 1999) whereas $\langle \Delta{\mbox{\rm
(B-R)}}$/$\Delta(\log\mbox r) \rangle$=$-0.09\pm 0.07$ is cited for
non-radio emitting galaxies (Peletier et al. 1990).  Therefore the
colour gradient of J0321+3410 is consistent with the above
trends, but perhaps a more extreme example. \\
\noindent There is a colour reversal, a blue-ward swing, at radial distances
r(arcsec)$<$3, clearly visible in the colour image and profile,
Figure~\ref{bmrprofile}. The radial colour gradient is $
\Delta{\mbox{\rm(B-R)}}$/$\Delta(\log\mbox r)$=$0.3\pm
0.02$. Since we are dealing with an AGN the obvious interpretation is
that the blue emission is AGN light. This is the extra light that
reduces the 4000 \AA\ break contrast below that expected for pure
starlight.  But it is important to note that this blue emission is not
point-like as the nuclear region becomes bluer at a radius $\sim$ 2-3
arcsec. Thus it is possible that a circumnuclear starbust component
exists near the AGN. We have also computed the ratio between the effective radii
r$_e$(B)$/$r$_e$(R), following Mahabal et al (1999), and we obtained
r$_e$(B)$/$r$_e$(R)=0.69, i.e. the central surface brightness in B is
larger than that of R. That is consistent with the observed colour gradient and
with Mahabal et al (1999) findings for a sample of radio galaxies
$\langle{\mbox{r$_e$(B)$/$r$_e$(R)}}\rangle =0.87\pm 0.15$, in
contrast with early-type galaxies
$\langle{\mbox{r$_e$(B)$/$r$_e$(R)}}\rangle=1.25\pm 0.1$.  
\begin{figure*}
\includegraphics[width=8.5cm]{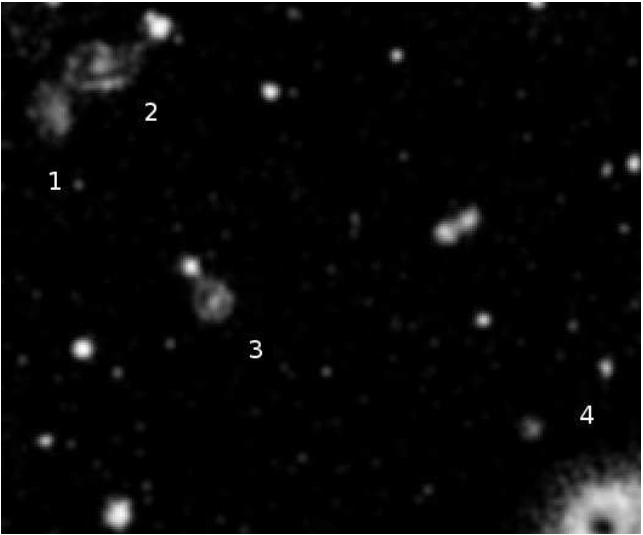}
\caption{B-R colour image of the galaxies (labeled as 1,2,3) in the 
field of view of 0324+3410 (labeled as 4). Field of view is $1.7'\times 1.3'$ }
\label{bmr}
\end{figure*}
\noindent There is a bluish patch in J0324+3410, at a distance of 4-5$''$  from the 
centre which shows up as a dip in the radial profile and a dark arc in B-R image. 
 That patch coincides with the ring region. Blue dips are observed in B-R profiles 
of galaxies with strong star formation (see figure 22 in Barton et al. 2003) and 
in ring galaxies, e.g. Arp 10 (see figure 20 in Appleton et
 al. 1997).  The interpretation is that the blue dips are regions of
 star-formation.  If we interpret the bluish patch in J0324+3410 as
 composed by clumps of young blue stars, it suggests triggering by
 interactions/collisions and mergers.  \\
\noindent  Is there independent evidence for an interaction in
J0324+3410?  Examination of the field surrounding J0324+3410, see
Figure~\ref{br} and Figure~\ref{bmr}, shows a pair of interacting
galaxies in the northeast (labeled as ``1'' and ``2'' in
Figure~\ref{bmr})\footnote{no redshifts are available}. They have a
disrupted morphology seen in B and R images, and are also both bluish
in colour.  There is a third galaxy (labeled as ``3'' in
Figure~\ref{bmr}) , between the pair of galaxies and J0324+3410, that shows a colour
gradient, being bluer in the region that is nearest to the pair of
galaxies. If this group of galaxies lies at the same redshift as our
object then the pair would be at a projected distance of approximately
145 kpc. Considering a typical relative velocity of 300 km$s^{-1}$
that would mean that the galaxies could have been interacting in the
last $\sim$ 10$^8$ years. Given that the age of star forming clumps in
interacting and merging systems range between few mega-years and
10$^8$ yrs (see Hancock et al. 2007 and references in-there), this
time scale is consistent with the possibility that the ring emission
comes from an interaction-induced starburst in J0324+3410.
 We have mapped  L-band VLA data from the archive, 
Figure~\ref{vla}. There is a core plus a two-sided structure, an unusual 
morphology for a blazar, suggesting that the AGN axis makes a larger
angle with the line of sight than is typical (i.e. $<$30$^o$). Note that this is 
consistent with the models that argue that NLS1 are objects seen at  large angle.  
A lower limit on the age of the source might be estimated through the radio size plus an assumption
on the lobe advance speed. The radio (unprojected ) size is $\sim$15 kpc, assuming a
 typical speed of  0.01c, that gives an age between 10$^7$- 10$^8$ years, for angles 
with the line of sight between 30-80$^o$. It is interesting the consistency between
the estimates of the time scales of interaction and radio activity, suggesting a correlation between the two, but note the caveat of the lack of redshift for the field galaxies. \\
\begin{figure*}
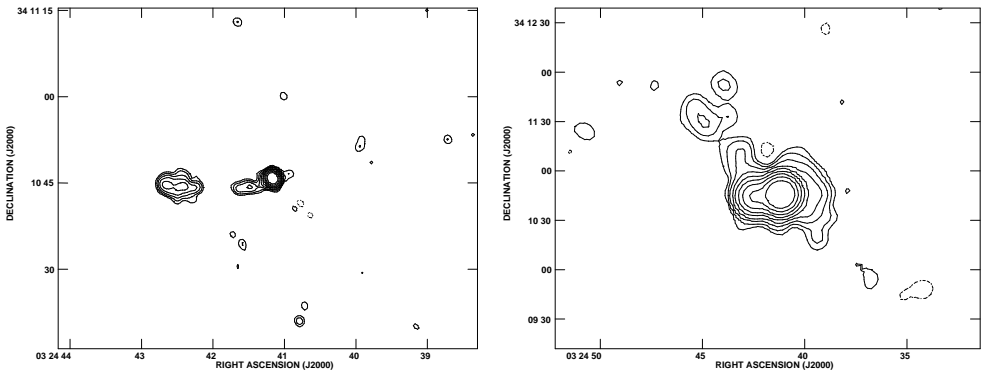

\includegraphics[width=5cm,angle=-90]{8926f6a.ps}
\includegraphics[width=5cm,angle=-90]{8926f6b.ps}
\caption{VLA 1.4 GHz maps of 0324+3410 in A (left) and C (right) configurations. The contours
increase in factors of 2 from 0.0004 Jy/beam for the A-configuration and 0.0005 Jy/beam for the C-configuration.}
\label{vla}
\end{figure*}
\noindent On the other hand, and by analogy with Arp 10 (CA96), J0324+3410 
might be a merger  remnant. 
 As in Arp 10, J0324+3410 shows a
faint outer asymmetric disk-like  emission. The colour gradient magnitude and
the colour of the structure inside and across the ring-morphology of
J0324+3410 is very similar to that see in collisional ring galaxies,
in particular to that of Arp 10. The later may be in the intermediate
stage of a merger , and this could
also be the case of J0324+3410, a minor merger involving a small
companion which has already been absorbed into the core of the central
object and leading to a ring-like structure.

\section{Summary}
We have combine B and R NOT images of J0324+3410 to study its host
galaxy. Analysis of the images shows a ring structure that resembles
the inner region of interacting/merger systems. The B-R colour image and B-R
profile of J0324+3410 reveal that: a) the object is bluer towards its
centre, and we interpret it as emission from the AGN plus
circumnuclear starburst b) the ring contains blue regions interpreted
as clumps of star forming regions c) there is a large colour gradient,
that is consistent with those detected in collisional ring galaxies. In
many aspects J0324+3410 is similar to the ring galaxy Arp 10.
We discuss scenarios in which J0324+3410 is a merger remnant, or
part of an interacting system in the last 10$^8$ years.

\section{Acknowledgments}
We thank the referee for constructive comments, and Chien Peng support with the GALFIT software.
Sonia Ant\'on acknowledges the financial support from the
Portuguese Funda\c c\~ao para a Ci\^encia e Tecnologia through the project 
ESO/FNU/43803/2001, SFRH/BPD/20859/2004 and from Jodrell Bank Observatory visitor grant.
 This research has made use of the NASA/IPAC Extragalactic
Database (NED) which is operated by the Jet Propulsion Laboratory,
California Institute of Technology, under contract with the National
Aeronautics and Space Administration. The NOT is operated on the island of La Palma
jointly by Denmark, Finland, Iceland, Norway, and Sweden, in the
Spanish Observatorio del Roque de los Muchachos of the Instituto de
Astrofisica de Canarias. The National Radio Astronomy Observatory is a facility 
of the National Science Foundation operated under cooperative agreement by Associated Universities, Inc.

\end{document}